\documentclass[letterpaper]{article} 
\usepackage{aaai24}  
\usepackage{times}  
\usepackage{helvet}  
\usepackage{courier}  
\usepackage[hyphens]{url}  
\usepackage{graphicx} 
\usepackage{natbib}  
\usepackage{caption} 
\usepackage{algorithm}
\usepackage{algorithmic}
\usepackage{multirow}
\usepackage{newfloat}
\usepackage{listings}
\usepackage{array,etoolbox}
\usepackage{caption} 
\usepackage{amsmath,graphicx}
\usepackage{tikz}
\usepackage{multirow}
\usepackage{amsmath}
\usepackage{bbding}
\preto\tabular{\setcounter{magicrownumbers}{0}}
\newcounter{magicrownumbers}
\newcommand\rownumber{\stepcounter{magicrownumbers}\arabic{magicrownumbers}}

\def\checkmark{\tikz\fill[scale=0.4](0,.35) -- (.25,0) -- (1,.7) -- (.25,.15) -- cycle;} 

\usepackage{algorithm}
\usepackage{algorithmic}

\usepackage{newfloat}
\usepackage{listings}
\DeclareCaptionStyle{ruled}{labelfont=normalfont,labelsep=colon,strut=off} 
\floatstyle{ruled}
\newfloat{listing}{tb}{lst}{}
\floatname{listing}{Listing}

\title{AAAI Press Formatting Instructions \\for Authors Using \LaTeX{} --- A Guide}

\title{Kid-Whisper: Towards Bridging the Performance Gap in Automatic Speech Recognition for Children VS. Adults}
\author {
    Ahmed Adel Attia\textsuperscript{\rm 1},
    Jing Liu\textsuperscript{\rm 1},
    Wei Ai\textsuperscript{\rm 1},
    Dorottya Demszky\textsuperscript{\rm 2},
    Carol Espy-Wilson\textsuperscript{\rm 1}
    }
\affiliations {
    \textsuperscript{\rm 1}University of Maryland\\
    \textsuperscript{\rm 2}Stanford University\\
    aadel@umd.edu, 	jliu28@umd.edu, aiwei@umd.edu, ddemszky@stanford.edu, espy@umd.edu	
}

\usepackage{bibentry}
\begin{document}

\maketitle
\begin{abstract}
    
Recent advancements in Automatic Speech Recognition (ASR) systems, exemplified by Whisper, have demonstrated the potential of these systems to approach human-level performance given sufficient data. However, this progress doesn't readily extend to ASR for children due to the limited availability of suitable child-specific databases and the distinct characteristics of children's speech. A recent study investigated leveraging the My Science Tutor (MyST) children's speech corpus to enhance Whisper's performance in recognizing children's speech. They were able to demonstrate some improvement on a limited testset. This paper builds on these findings by enhancing the utility of the MyST dataset through more efficient data preprocessing. We reduce the Word Error Rate (WER) on the MyST testset 13.93\% to 9.11\% with Whisper-Small and from 13.23\% to 8.61\% with Whisper-Medium and show that this improvement can be generalized to unseen datasets. We also highlight important challenges towards improving children's ASR performance and the effect of fine-tuning in improving the transcription of disfluent speech.  
\end{abstract}
\section{Introduction}
\label{intro}
\label{sec:intro}
Automatic Speech Recognition (ASR) has witnessed a boom in recent years through utilizing huge amounts of transcribed speech scraped from the internet. Whisper \cite{radford2023robust} was able to approach human-level accuracy by utilizing 680K hours of speech data. In contrast, XLS-R \cite{babu2021xls} pre-trains on 436K hours of untranscribed speech in a self-supervised manner and 65K hours of transcribed speech. Both models were able to achieve state-of-the-art (SOTA) results by leveraging huge amounts of data. ASR models still underperform with low-resource languages and tasks. Recent works have attempted to explore how ASR models performance can be improved for low-resource languages \cite{talafha2023n, 9053139, yi2020applying, reitmaier2022opportunities} but they haven't caught up with high-resource languages. 

Children ASR is considered a low resource task and previous works have demonstrated the gap between children and adult ASR even in English. The main reason for that gap has been attributed to inter-speaker variability due to varying developmental rates and intra-speaker variability due to underdeveloped pronunciation skills \cite{koenig2008speech, koenig2008stop, lee1999acoustics, lee1997analysis, vorperian2007vowel,  smith1992relationships}. Current ASR models trained on adult speech are not capable of learning these variabilities as they are mostly unseen in the training data. Moreover, children's speech databases are limited and difficult to collect and transcribe \cite{claus2013survey}.

In this work, we explore how Whisper can be fine-tuned on children's speech. We chose Whisper because of its massive training data which makes it more likely to generalize to unseen and uncommon speech patterns. Additionally, Whisper has been shown to be noise-robust \cite{gong2023whisper}. We take advantage of the My Science Tutor (MyST) speech corpus \cite{ward2019my} which is the largest publicly available children's speech corpus, provided free to academics for research purposes.

A recent study \cite{jain2023adaptation} has attempted to adapt Whisper to the MyST corpus. They found that the quality of audio files as well as transcriptions in the MyST corpus varies, and were able to extract 65 hours of well-transcribed speech from the 197 hours of transcribed speech provided in MyST. We expand upon their work by outlining a more efficient data preprocessing scheme and extracting a total of 179.2 hours, which we show improves the performance and robustness of Whisper. Additionally, we maintain the train/test/development splits provided in the MyST corpus to ensure there's no overlap in speakers between data splits. We demonstrate tangible improvement on the MyST testset, reducing the Word Error Rate (WER) of the Small Whisper model from 13.93\% to 9.11\% and that of the Medium model from 13.23\% to 8.61\%. This also leads to improving the WER on the spontaneous part of the CSLU Kids dataset from 32.00\% to 27.16\% with the Small model, and from 31.85\% to 16.53\% with the Medium model without explicitly including this dataset in the training set.\\

Our contributions in this research are threefold
\begin{itemize}
    \item A systematic study of the effectiveness of transfer learning in adapting large ASR models to children's speech in a generalizable way.
    \item Our pre-trained models achieve SOTA performance on children's datasets, both seen during training (MyST and CSLU scripted) as well as completely unseen during training (CSLU Spontaneous) without harming their performance on adult speech (Librispeech test-clean).
    \item Our preprocessing technique was critical in improving the usability of a publicly available dataset that other works haven’t fully utilized. The data used in our fine-tuning of Whisper is the largest in children ASR to date.
    \item Detailed error analysis, which highlights our model's improved ability to capture disfluencies and to handle unusual grammatical structures present in children's speech as well as current shortcomings and challenges in children's ASR.
 
\end{itemize}

We begin by giving a quick overview of Whisper in Section 2, followed by a description of the datasets used and our proposed preprocessing scheme in Section 3. We follow that by showcasing our experiments and training parameters in Section 4. Results are in Section 5 followed by detailed error analysis in Section 6. We end with a conclusion, and an outline of our future research plans in Section 7.

\section{Model Description}
\label{sec:Model}

Whisper is a family of ASR models with varying sizes, namely, Tiny, Base, Small, Medium, and Large. Models from Tiny to Medium have an English-only variant and a multilingual variant. The training data for Whisper includes 438K hours of English-to-English transcription, 117K hours covering 96 languages not including English, and 125K hours of speech spoken in different languages, transcribed in English. The English-only variants of Tiny to Medium were only trained on the 438K hours of English-to-English transcription. To filter out low-quality transcription, the training set was passed through an initial model, and files with a high WER were flagged and manually inspected to remove automatically transcribed and mistranscribed files. The remaining substantial amount of training data helped Whisper achieve near human-level transcription, especially in English, with their Large model achieving a WER of 2.82 on the Librispeech clean test set.

\section{Dataset Description and Processing}
\label{sec:dataset}
We mainly focus on the MyST corpus in this study. However, we also discuss how well the results on MyST can be generalized beyond this corpus. For that purpose, we use the scripted and spontaneous CSLU kids databases \cite{shobaki2000ogi}. Finally, we study how finetuning affects the performance on adult speech by testing our models on the test-clean subset of Librispeech.  In this section, we describe each corpus. 
\subsection{My Science Tutor Dataset}
The MyST corpus is the largest publicly available children's speech corpus. It consists of 393 hours of conversational children's speech, recorded from virtual tutoring sessions in physics, geography, biology, and other topics. The corpus spans 1,371 third, fourth, and fifth-grade students although age and gender information for each student are not available. Around 197 hours of the dataset were transcribed, although the quality of transcriptions varies. To the best of our knowledge, the MyST corpus was not included in Whisper's training set. 
Upon manual inspection, some transcriptions were assigned to the wrong files completely.
\begin{quote}
    Provided Transcription: Um, the wires are like a pathway energy goes through it into the motor and makes it work.\\
    Actual Transcription: Um, because it's metal, and metal I think has energy.
\end{quote}
Other files appear to have been automatically transcribed with a lower-quality automatic transcriber.
\begin{quote}
    Provided Transcription: No, I don't hearing even a candle burns.\\
    Actual Transcription: No, I don't hear anything when the candle burns.
\end{quote}
Additionally, some files have poor audio quality, with the children speaking too close to the microphone, which resulted in a high level of distortion in the audio files.

To identify these files, we follow a similar technique as in \cite{radford2023robust}, by passing the entire dataset through Whisper-Large and flagging files with WER larger than 50\%. Additionally, one and two-word files were removed altogether, because they lacked the context to distinguish between homophones, like "to", "too" and "two". All files with no speech activity, i.e. files labeled as $<$DISCARD$>$ or $<$NO\_SIGNAL$>$ or $<$SILENCE$>$, were also removed from the dataset.  Table \ref{table: filtering} shows the effect of different filtering steps on total dataset duration and WER. According to our results, around 5 hours of the training data is either mistranscribed or has low audio quality and is responsible for increasing the WER on the training data by about 3\%. Similar results can be inferred about the test and development sets. Additionally, short files which accounted for only 4 hours of the training data increased the WER by more than 7\%. We will publish the list of flagged files on GitHub and link to it in the camera-ready manuscript. 
\begin{table}[h]

  \caption{WER of Whisper-Large-v1 transcriptions of all three data splits of the MyST corpus before and after different levels of filtration (Duration of splits in hours).}
  \centering
  \resizebox{\columnwidth}{!}{
  \Large
  \begin{tabular}{|c |c | c| c|}
    \hline\rule{0pt}{3ex}
   \textbf{Filteration Method} & \textbf{Train} &\textbf{Test} & \textbf{Development} 
    \\[1pt]\hline\rule{0pt}{3ex}    
     \textbf{No Filteration} & 29.5 (145) & 26.2 (28.1) & 26.2 (25.5)\\
    \hline 
    \textbf{Removing Files w. WER $>$ 50\%} & 26.8 (140) & 22.3 (26.7) & 22.3 (25.5)\\
    \hline
    \multicolumn{1}{|c|}{\begin{tabular}{@{}c@{}}\textbf{Removing Files w. WER $>$  50\%} \\ 
   \textbf{or w. Less Than 3 Words}\end{tabular}} & 19.2 (132.5) & 14.2 (25.6) & 12.8 (21)
    \label{table: filtering}
    \\\hline
  \end{tabular}}
\end{table}

\begin{table}[]

  \caption{Summary of the Datasets Used. Duration in hours.}
  \centering
  \resizebox{\columnwidth}{!}{
  \Large
  \begin{tabular}{|c |c| c| c| c| c|}
    \hline
   \rule{0pt}{4ex}\textbf{{Dataset}} & \multicolumn{1}{c|}{\begin{tabular}{@{}c@{}}\textbf{{Training}} \\ \textbf{{Duration}}\end{tabular}} &\multicolumn{1}{c|}{\begin{tabular}{@{}c@{}}\textbf{{Development}} \\ \textbf{{Duration}}\end{tabular}} &\multicolumn{1}{c|}{\begin{tabular}{@{}c@{}}\textbf{{Testing}} \\ \textbf{{Duration}}\end{tabular}} &\textbf{{Filtered?}} &\textbf{Age Group}
    \\[5pt]\hline
    \rule{0pt}{2ex}
     \textbf{MyST} & 125  & 20.9  & 25.8  & \checkmark & 8-11 Years\\
     \textbf{CSLU Kids - Scripted} & 35  & 4.8  & 4.8 & \text{\sffamily X}  & 6-11 Years \\
     \textbf{Librespeech- testclean} & 0 & 0 & 5.4  & \text{\sffamily X} & Adult
    \label{table: dataset_calc}
    \\\hline
  \end{tabular}
  }
\end{table}

Files longer than 30 seconds in the training and development sets were also removed. That is because Whisper processes files in 30-second chunks, and any files longer than 30 seconds are truncated. However, it is not possible to accurately truncate the transcriptions without any timestamps present, so these files are unsuitable for loss calculation. Additionally, the majority of the files in the MyST corpus were too short, with the average file length in the training data being 8 seconds. That would mean that training batches are mostly padding, leading to inefficient training. To remedy this problem, files within a single recording session were concatenated to be close to but not longer than 30 seconds while maintaining the context of the conversation within the recording session. 

Our filtering technique removes 17.8 hours from the entire dataset, which leaves us with 179.2 hours of well-transcribed speech in total. We maintain the train/development/test split provided in the MyST database to avoid any overlap in speakers between the splits. We ended up with 132.5 hours in the training data, 20.9 hours in the development data, and 25.8 hours in the test data.

The text of the transcriptions was all upper case which destabilized the training. Consequently, all the text was mapped to be lowercase and further normalized using  the WhisperNormalizer\footnote{https://pypi.org/project/whisper-normalizer/} Python package, which mapped tokens like "you're" to a standard "you are", as well as mapping all digit numbers to be spelled out. This ensured that only actual mistranscriptions would be penalized. This also reduces the diversity in transcription quality, which was noted to harm the performance, unlike diversity in audio quality\cite{radford2023robust}.

In contrast to our filteration system, the method used by \cite{jain2023wav2vec2} resulted in only 65 hours of data. This data was partitioned into a 55-hour training set and a 10-hour test set and no development set. Their sets also suffered from overlapping speakers between the train and test sets. By sticking to the splits provided by MyST, our splits share no overlapping speakers, and have almost 3x the data. 
\subsection{CSLU Kids}
The CSLU Kids speech corpus contains spontaneous and prompted speech from 1100 children between Kindergarten and Grade 10, with approximately 100 children per grade. In the scripted subset of the dataset, each child was prompted to read from a list of 319 scripts, which can either be simple words, sentences, or digit strings.
Each utterance of spontaneous speech begins with a recitation of the alphabet followed by one minute of unprompted speech. The spontaneous speech in the CLSU corpus is distinct from the MyST corpus in that it is unstructured. Instead of talking about a particular topic, children were only given an open prompt like "Tell me about your favorite movie." \cite{shobaki2000ogi}. Below is a transcription sample.
\begin{quote}
    ...usually just lay down on my bed, for now i don't like to i don't know, uh football okay first they are like standing on the ground and then they run and then they mm and if the girl pass the whole field you get a six points uh think it's twenty four i don't know think yeah they catch block and uh one uh the quarter back throws and the runners run uh it's blue uh and it has a big big big electric train set uh i have a workshop...
\end{quote}

Notice how this single utterance talks about multiple topics, which is typical in children's speech and provides a unique challenge to the decoder in Whisper which acts as an audio-conditional language model which relies on linguistic context as well as audio representations in its output.

The majority of the recordings in the spontaneous section of the CLSU corpus were longer than 30 seconds, and are thus unsuitable for training. Instead, we use the scripted portion of the CSLU corpus to help the model adapt to the channel differences between MyST and CSLU recordings but still consider the spontaneous section as an out-of-sample testset.

The transcriptions were of a high enough quality and filtering was not necessary, but they were all normalized to ensure a standard transcription style. Files in the scripted portion of the dataset were shuffled and split into train, development, and test sets with an 80/10/10 split. The training set was 35 hours long, and the development and test sets were both around 4.8 hours long. Short files were combined to be close to 30 seconds as we did with the MyST corpus. 

\subsection{Librespeech: test-clean}
The test-clean subset of the Librespeech corpus was used to test the ASR model's performance on Adult speech. It contains about 5.4 hours of speech read from Audiobooks from the LibriVox project. Since Librespeech was not used for training, we didn't combine the files, and we also didn't filter out any transcriptions to allow for reproducible and contrastable results. All transcriptions were normalized.

\section{Training Details and Hyperparameters}

We followed the Huggingface Whipser finetuning tutorial \footnote{https://huggingface.co/blog/fine-tune-whisper}. Our evaluation script, which calculates the WER, was adapted from a code snippet by OpenAI\footnote{https://github.com/openai/whisper/discussions/654}. All models were trained on Nvidia A6000 50GB GPU.

For the Small models, we used a learning rate of $1\times 10^{-5}$, 1 gradient accumulation step, and a batch size of 64 for the English-only model, and 32 for the multilingual model.  For the Medium models, we used a learning rate of $1\times 10^{-5}$, 1 gradient accumulation step, and, a batch size of 32 for both the English-only and multilingual models. All models were finetuned until convergence and the best checkpoints were used for evaluation. 

\section{Results and Discussion}
\subsection{Whisper Zero-shot Models}
Table \ref{table: outhebox} shows the WER for different Whisper models without any finetuning. Looking at these results, the gap between children and adult speech becomes immediately clear.  The WER for the scripted part of CSLU Kids is between 6 and 10 times that of Librispeech, and the WER for MyST is between 3 and 5 times.  In general, English models perform better than multi-lingual models, with the exception of the Medium model where the results are mixed. That could be because the Medium model is big enough to benefit from seeing more data in different languages. The bigger the model, the better the performance, with the exception of Large-V1 being worse than Medium. In fact, the performance seems to saturate beyond Medium and the difference in performance between Medium and Large-V2 is negligible.\\
We note that the zero-shot WER reported here is smaller than that reported in \cite{jain2023adaptation}. We attribute this to the fact that they used a different normalizer than the one Whisper was trained with, which we validated by inspecting their datasets which are publicly accessible on Huggingface.
Since the performance saturates beyond Medium, we finetune the Small and Medium models, both the English and multilingual variants. 
\begin{table}[]

  \caption{Zero-shot WER on different test sets for different Whisper Models Without Finetuning.}
  
  \centering
  
  \resizebox{\columnwidth}{!}{
  \begin{tabular}{|c |c|c| c|| c|}
    \hline
   \rule{0pt}{4ex}
   \textbf{Model} & \textbf{MyST} &\multicolumn{1}{c|}{\begin{tabular}{@{}c@{}}\textbf{CSLU Kids} \\ 
   \textbf{Scripted}\end{tabular}}  &\multicolumn{1}{c||}{\begin{tabular}{@{}c@{}}\textbf{CSLU Kids} \\ 
   \textbf{Spontaneous}\end{tabular}}  &\multicolumn{1}{c|}{\begin{tabular}{@{}c@{}}\textbf{Librespeech} \\ 
   \textbf{testclean}\end{tabular}} 
    \\[1pt]\hline\rule{0pt}{3ex}    
     \textbf{Tiny} & 21.16 &	74.98 & 57.01	& 7.49 \\
     \textbf{Tiny.en} & 18.34 & 61.04  & 45.29 & 5.59\\ 
     \textbf{ Base} & 18.54 & 40.20 & 43.71 & 4.98\\
     \textbf{Base.en} & 15.57 & 33.18 & 38.57 & 4.15 \\
     \textbf{Small} & 14.06 & 25.15 & 36.36 & 3.39\\
    \textbf{Small.en} & 13.93 & 21.31 & 32.00 & 3.05\\
    \textbf{Medium} & 12.90 & 18.62 &37.04&2.76\\
    \textbf{Medium.en} & 13.23 &18.57 &31.85&3.02\\
    \textbf{Large-V1} & 14.15 &21.50 &45.18&2.98\\
    \textbf{Large-V2} & 12.80 &17.22 &29.39&2.82    
    \label{table: outhebox}
    \\\hline
  \end{tabular}}
\end{table}
\subsection{Finetuned Whipser Models}
In this section, we showcase the performance of our finetuned models and contrast them with the models from \cite{jain2023adaptation}, whose models are publicly available on Huggingface. We report the best-performing variants here. We tested all models on testsets from four corpora, listed in Table \ref{table: dataset_calc}.

\begin{table*}[t]

  \caption{WER on different test sets for different Whisper Models. EN stands for English-only model and ML stands for multilingual model. }
  \centering
  \resizebox{2\columnwidth}{!}{
  \begin{tabular}{|c |c|c| c| c || c|}
    \hline
   \rule{0pt}{4ex}
   \textbf{Model} &\textbf{Training Data}&\textbf{MyST} &\multicolumn{1}{c|}{\begin{tabular}{@{}c@{}}\textbf{CSLU Kids} \\ 
   \textbf{Scripted}\end{tabular}}  &\multicolumn{1}{c||}{\begin{tabular}{@{}c@{}}\textbf{CSLU Kids} \\ 
   \textbf{Spontaneous}\end{tabular}}  &\multicolumn{1}{c|}{\begin{tabular}{@{}c@{}}\textbf{Librespeech} \\ 
   \textbf{testclean}\end{tabular}} 
    \\[1pt]\hline
    \multicolumn{6}{|c|}{\textbf{Small}}\\
    \hline
     \rule{0pt}{3ex}    \textbf{ML - zero-shot} & - &14.06 &25.15 &36.36 &3.39 \\
     \textbf{EN - zero-shot} &- &13.93  &21.31  &32.00 &\textbf{3.05} \\
     \hline\rule{0pt}{3ex}
    \textbf{EN - \cite{jain2023adaptation}} & MyST55H&13.23 &31.26 &28.63&5.40\\
     \textbf{ML} & MyST &11.80  &55.51 &28.53 &6.23 \\ 
     \textbf{ML} & MyST + CSLU &12.11  &2.74 &32.72 &7.97 \\ 
    \textbf{EN} & MyST &\textbf{9.11}  &33.85 &28.47 &\textbf{4.18} \\ 
     \textbf{EN} &  MyST + CSLU&9.21  &\textbf{2.59} &\textbf{27.16 } &4.74 \\ 
    \hline
    \hline
       \multicolumn{6}{|c|}{\textbf{Medium}}\\
    \hline
  \rule{0pt}{3ex}
      \textbf{ML - zero-shot} & - &12.90 &18.62 &37.04 & \textbf{2.76}\\
\textbf{EN - zero-shot} & - &13.23 &18.57 &31.85 & 3.02\\
    \hline \rule{0pt}{3ex}
    \textbf{EN - \cite{jain2023adaptation}} &  MyST55H&14.40 &28.31 &26.76&8.66  \\
    \textbf{ML} & MyST  &\textbf{8.61} &30.10 &24.26 &5.32 \\ 
     \textbf{ML} & MyST + CSLU &8.99 &\textbf{1.97} &20.28&4.28  \\
    \textbf{EN}& MyST &8.91  &47.94  &25.56 &3.95 \\ 
     \textbf{EN} &  MyST + CSLU &8.85  &2.38  &\textbf{16.53} & \textbf{3.52}\\
    \hline
  \end{tabular}}
    \label{table: results}
\end{table*}

Looking at the results in Table \ref{table: results}, it is clear that Whisper can and does improve its performance on the MyST dataset as well as CSLU, proving that transformer ASR models have the capacity to improve their performance on children's speech. We establish strong SOTA performance of 8.61\% for the MyST testset. Our best performance on the CSLU scripted dataset of 1.97\% beats the current SOTA of 5.52\% \cite{yeung2021fundamental} which is the current SOTA to the best of our knowledge. We also show improvement on unseen datasets, since our models trained on just MyST, or a combination of MyST and CSLU scripted data show improvement on CSLU Spontaneous speech without any training on speech from this dataset. Our best-performing model on the CSLU Spontaneous dataset scores 16.53\% WER, which is about half the WER of zero-shot Whisper. Additionally, our models ``forget'' less about the adult speech than the baseline, with our models seeing a degradation of only about 1\%.

Medium models outperformed Small models, and generalized better to unseen datasets. The English-only variant of the Small model showed significant improvement over the multilingual variant in seen and unseen datasets. The Medium multilingual variant performed slightly better on the MyST dataset when finetuned exclusively on it, but the English-only variant generalized better to unseen data. Multilingual models in both sizes had higher WER for Librispeech.

Looking at the results for the scripted portion of the CSLU corpus, it is clear that the lack of context in these scripts harm the performance of the models that weren't trained on speech from this dataset. However, the performance improved significantly when speech from this dataset was included in the training data, mainly because of the lack of variability of the scripts, unlike the more diverse MyST or CSLU Spontaneous datasets. We also attribute the gap in performance between the MyST and CSLU Spontaneous datasets to the fact that speech in the MyST corpus is more structured than the CLSU Spontaneous dataset. This shows that one of the reasons behind the gap in performance between adult and children's ASR is that the decoder in Whisper, which acts as an audio-condtional language model, is not well adapted to the variablility found in children's speech, where they can suddenly change topic several times in a short period. We discuss the role of the audio-conditional language model in handling disfluencies in children's speech in the next section.

\section{Error Analysis}
In this section, we analyze some example transcriptions from the Zero-shot Medium.en model as well as the Medium.en model trained on MyST and CSLU Scripted to gain insight into how fine-tuning Whisper can improve the performance on different domains. Table \ref{tab:examples} shows such examples.\\
\begin{table*}[t]
  \centering
  \begin{tabular}{|c |c |c |}
    \hline
    ID & Model & Transcription \\
    \hline 
    \multicolumn{3}{|c|}{\textbf{MyST}}\\
    \hline
    \multirow{3}{*}{\textbf{\rownumber}} & Ground Truth & Makes like the thing turn on and gets like a magnet so the little washers will stick \\

    & Medium.en Fine-tuned & makes like the thing turn on and gives like a magnet it the little washers will stick\\
    & Medium.en zero-shot & makes the thing turn on it gives a magnet the little washers will stick \\
    \hline
    \multirow{2}{*}{\textbf{\rownumber}} & Ground Truth & the water is plain and the grape it the grape tastic thing is plain the salt\\
    & Medium.en Fine-tuned & the water is plain and the grape is the grape tastic thing is plain the salt \\
    & Medium.en zero-shot & the water is plain and the grape the grape tastic thing is plain dissolved \\
    \hline
    \multirow{2}{*}{\textbf{\rownumber}} & Ground Truth & and it is pretty much into m turned into it is pretty much turned into a magnet \\
    & Medium.en Fine-tuned & and it is pretty much into ma turned into it is pretty much turned into a magnet\\
    & Medium.en zero-shot & and it is pretty much turned into a magnet \\
    \hline
    \hline
    \hline 
    \multicolumn{3}{|c|}{\textbf{CSLU Spontaneous}}\\
    \hline
    \multirow{3}{*}{\textbf{\rownumber}} & Ground Truth & a b c d e f g h j i k l m n o p q r s t u v w x y and z uhm he is like a grumpy man \\
    & Medium.en Fine-tuned & he is like a grumpy man\\
    & Medium.en zero-shot & he is like a grumpy man \\
    \hline
    \multirow{3}{*}{\textbf{\rownumber}} & Ground Truth & uhm i have the lion king toys uhm no uhm i forgot the babies have a pink dress on \\
    & Medium.en Fine-tuned & i have the lion king toys no i forgot my babies have a pink dress on\\
    & Medium.en zero-shot & it has a lion king toys a pink dress on \\
    \hline
   \multirow{3}{*}{\textbf{\rownumber}} & Ground Truth & ...q r s t u v w x y z tar i do not know just rock i guess pretty strong probably making music  \\
    & Medium.en Fine-tuned & ...q r s t u v w x y z i do not know just rock i guess pretty strong probably making music \\
    & Medium.en zero-shot & yeah probably making music\\
    \hline
  \end{tabular}
  \caption{Example transcriptions from the Medium.en model trained on MyST + CSLU Scripted and the zero-shot Medium.en model along with ground truths from the MyST and the CSLU Spontaneous. All the transcriptions were normalized using the WhisperNormalizer package.}
  \label{tab:examples}
\end{table*}

\subsection{MyST}
The examples from the top half of the table exemplify the main modes of error made by both our fine-tuned model and the zero-shot Whisper model.\\

Example 1 shows the main improvement of our model in comparison to the zero-shot model, which  is filler word and disfluency transcription. Whisper is known to perform implicit disfluency removal as a result of its data normalization \cite{radford2023robust}. Such filler words are common in children's speech \cite{gosy2019filler} and can provide useful insight into the development of children's speech. In this example, our model learned not to skip the filler word "\textit{like}" but still made a mistake with the word "\textit{gets}" which it transcribed as "\textit{gives}". This gives insight that these errors are caused by the audio-conditional language model in Whisper's decoder, which still struggles with unusual or incorrect grammatical structures. Another example of this is Example 2, where the disfluency in "\textit{...the grape is, the grape tastic thing is...} is mistranscribed by both models. The zero-shot model completely skips the word "\textit{is}" while our model mistranscribes is as "\textit{it}". Our model correctly transcribes the word "\textit{the salt}" which is added on to the end of the sentence and does not belong to any sound grammatical structure, while the zero-shot model mistranscribes it as "dissolved" which makes more grammatical sense but is not what was said. Both of these examples show that the audio-conditional language model might provide some further performance improvement possibly by further fine-tuning of the decoder.\\

Example 3 shows another example of disfluent speech that our model was better able to handle than the zero-shot model. The disfluency in "\textit{...and it is pretty much into a m.. turned into a magnet}" where the child forgets the phrase "\textit{turned into}" and starts to say "\textit{magnet}" but corrects themself as they start to say it - is correctly captured by the finetuned model, but it transcribes the cut-off word as "\textit{ma}" instead of "\textit{m}".  The zero-shot model completely discards the disfluency. \\
\subsection{CSLU Spontaneous}
When examining example transcriptions from the CLSU Spontaneous dataset, it is important to note that while the model was trained on the CSLU Scripted dataset, it was not trained on any examples from the CSLU Spontaneous dataset. That means that while the model might have been conditioned to the acoustic properties in this dataset, it was not exposed to the linguistic properties in CSLU Spontaneous, which is unique from both CSLU Scripted (since it is not read speech) and MyST (since it is not structured spontaneous speech).\\

Example 4 shows the main source of WER in both our fine-tuned model and the zero-shot model's transcription, which is that sometimes our model completely skips the recitation of the alphabet at the beginning of every recording. This is not a consistent problem, since sometimes our model transcribes the recitation like example 6 and it generally skips it less often than the zero-shot model. However, we did not see a pattern in utterances where it skips versus utterances where it does not, and further analysis is needed. \\

Example 5 shows that like in MyST, fine-tuning improves Whisper's capacity to handle disfluencies in children's speech. While our model consistently skips "\textit{uhms}" which were mistakenly removed during normalization, the zero-shot transcription tries to completely remove the disfluency while also mistranscribing "\textit{i have}" to "\textit{it has}", and as a result, completely alters the sentence.\\

Example 6 shows an utterance where our model transcribes the alphabet recitation while the zero-shot model removes it and a large chunk of the sentence, starting the transcription more than 20 seconds into the recording.\\

\section{Conclusions and Future Work}

In this paper, we outlined how Whisper, a SOTA ASR system can be finetuned on children's speech using MyST, the largest publically available conversational children's speech corpus. We showcased a way to filter out mistranscribed files from the corpus and established a strong baseline for children's speech recognition. Our finetuning reduced the WER by 4 to 5\% and reduced the gap between adult and children's speech. We also outlined some of the challenges that faces children ASR, namely the fact that audio-condtional language models are not well adapted to the variability in children's speech. Our error analysis shows that fine-tuning improves Whisper's ability to handle disfluencies in children's speech, and the language model's ability to handle unusual grammatical structures in children's speech, although there is room for improvement in the latter.

In the future, we will explore the noise robustness of Whisper. Specifically we will look at babble noise and other typical classroom nonspeech sounds and how they can affect performance, and how to improve such robustness in children's ASR. We will also explore whether these models are biased toward a certain gender, racial group, or age group. 

The authors of \cite{yeung2018difficulties} developed grade-specific ASR models, and proposed grouping different age groups separately, instead of under the umbrella term "children speech". Their suggested grouping was kindergarten; 1st grade; 2nd and 3rd grade; and 4th grade and above, and they noted that it is possible to achieve adult-like performance with the latter group. We aim to expand upon their work in the future, exploring whether their results can be replicated with large transformer ASR models and whether such bias against youger children can be mitigated.

\section{Acknowledgments}
The authors of this paper thank Wayne Ward for sharing his experience with Whisper, MyST, and other children's databases.


\bibliography{main}

\end{document}